\documentclass[sigconf]{acmart}

\AtBeginDocument{%
  }

\acmISBN{978-1-4503-XXXX-X/2018/06}

\usepackage{booktabs}
\usepackage{multirow}
\usepackage{tabularx}
\usepackage{array}
\usepackage{algorithm}
\usepackage{algpseudocode}
\usepackage[most]{tcolorbox}
\usepackage{listings}
\usepackage{xcolor}
\usepackage{framed}

\newcommand{\paratitle}[1]{%
  \par\addvspace{1.5ex}%
  \noindent\textbf{#1}\enspace\ignorespaces
}

\newtcolorbox{promptbox}[1]{%
  enhanced jigsaw,
  breakable,
  title={#1},
  colback=black!1,
  colframe=black!20,
  colbacktitle=black!12,
  coltitle=black,
  fonttitle=\bfseries,
  fontupper=\small,
  boxrule=0.4pt,
  arc=2mm,
  left=2mm,
  right=2mm,
  top=1.5mm,
  bottom=1.5mm,
  before skip=5pt,
  after skip=6pt,
  before upper={%
    \setlength{\parindent}{0pt}%
    \setlength{\parskip}{2pt}%
  }
}
\begin{document}

\title{BOUND: Brief-Guided Corrective Preference Distillation at Search-Control Boundaries}

\author{Qingying Niu}
\email{qingyingniu36@gmail.com}
\affiliation{%
  \department[0]{Gaoling School of Artificial Intelligence}
  \institution{Renmin University of China}
  \city{Beijing}
  \country{China}
}

\author{Ruiyang Ren}
\authornote{Corresponding authors.}
\email{reyon\_ren@outlook.com}
\affiliation{%
  \department{Gaoling School of Artificial Intelligence}
  \institution{Renmin University of China}
  \city{Beijing}
  \country{China}
}

\author{Wayne Xin Zhao}
\authornotemark[1]
\email{batmanfly@gmail.com}
\affiliation{%
  \department{Gaoling School of Artificial Intelligence}
  \institution{Renmin University of China}
  \city{Beijing}
  \country{China}
}

\author{Yaliang Li}
\email{yaliang.li@alibaba-inc.com}
\affiliation{%
  \institution{Alibaba Group}
  \city{Bellevue}
  \state{Washington}
  \country{United States}
}

\renewcommand{\shortauthors}{Niu et al.}

\begin{abstract}
Large language model (LLM)-based deep search agents solve tasks through iterative retrieval and reasoning, but locally relevant evidence can cause persistent wrong-anchor drift, constraint drift, or local-topic drift. Existing methods supervise trajectories, outcomes, or steps, but rarely distinguish task-aligned continuations from locally plausible ones that reinforce drift. We propose BOUND, a brief-guided corrective preference distillation framework for persistent search drift. For each student-induced decision-time state, BOUND constructs a teacher-side search-state brief that preserves the original search target and key constraints while summarizing confirmed evidence, missing information, and drift status. Guided by the brief, the teacher determines whether the student's continuation contains a correctable local search-control error likely to affect subsequent decisions. Together with the rollout outcome, this assessment determines whether to construct a corrective contrast between a student-specific correction and the original continuation, or a termination contrast between a supported answer and an unnecessary retrieval continuation. Each validated state-matched preference pair operationalizes a search-control boundary. Direct preference optimization (DPO) distills these preferences into the student, while the brief and teacher-side computation remain confined to training. We evaluate BOUND on four multi-hop QA benchmarks and three deep-search benchmarks. Across the six benchmarks for which we reran baselines, BOUND leads on five datasets and 12 of 14 metrics. Under the same search-control interface and matched settings, BOUND outperforms Trajectory SFT by 5.6 EM points on Bamboogle and 4.8 accuracy points on BrowseComp-Plus. Code is available at \url{https://github.com/RUCAIBox/BOUND}.
\end{abstract}




\setcopyright{none}
\settopmatter{printacmref=false}
\maketitle

\section{Introduction}
\label{sec:introduction}

Large language models (LLMs) show strong capabilities in language understanding, generation, and reasoning~\citep{achiam2023gpt}. Despite these capabilities, answering questions that require current or verifiable information often depends on external evidence. Retrieval-augmented generation (RAG) grounds generation in retrieved documents~\citep{lewis2020retrieval}, but one retrieval step is insufficient when evidence must be progressively acquired and connected. LLM-based search agents address this through iterative querying, retrieval, and reasoning. Recent work advances this process through agentic retrieval~\citep{li2025searcho1} and search-oriented post-training~\citep{jin2025searchr1,song2025r1searcher,li2025websailor}. However, dependence on retrieval makes search vulnerable to error propagation, as each result reshapes context and later decisions.

\begin{figure}[t]
\centering
\includegraphics[width=\linewidth]{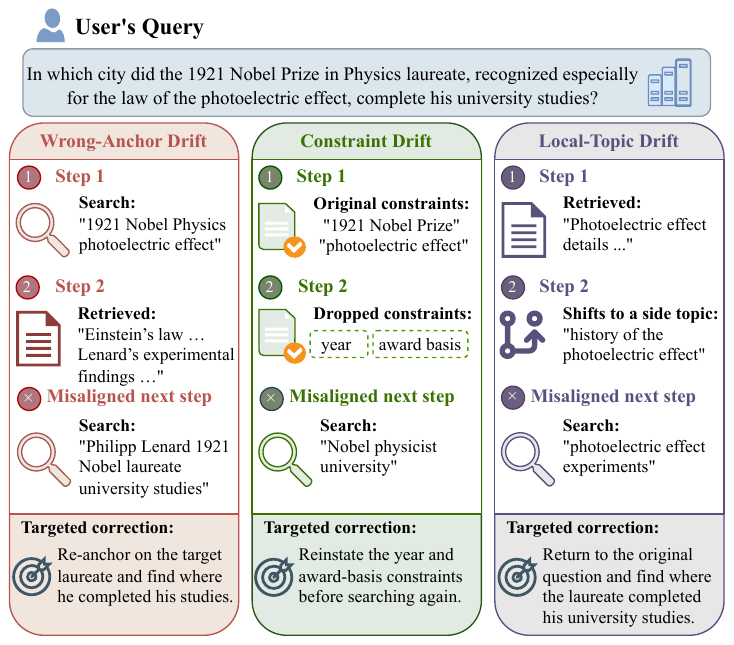}
\caption{Three manifestations of persistent search drift and their corrective search decisions.}
\Description{Illustration of wrong-anchor, constraint, and local-topic drift with the corresponding corrective search decisions.}
\label{fig:intro}
\end{figure}

Semantically relevant evidence can still mislead reasoning, allowing errors to propagate through multi-step search~\citep{dai2026llm}. Once the search departs from the original search target or drops a key constraint, later decisions may reinforce the deviation even when queries appear reasonable. We refer to this failure mode as \emph{persistent search drift}. Figure~\ref{fig:intro} illustrates three manifestations. In \emph{wrong-anchor drift}, a related entity introduced by retrieved evidence becomes the new search anchor, causing the agent to pursue Lenard instead of Einstein. In \emph{constraint drift}, the agent drops qualifiers concerning the year or the reason for the award. In \emph{local-topic drift}, it explores the history of the photoelectric effect rather than resolving the connection between the physicist's university studies and the target city. Across cases, the search remains locally plausible while diverging from the original search target.

Existing methods train search agents through trajectory-based post-training~\citep{li2025websailor,deng2026fort} and outcome-based reinforcement learning~\citep{jin2025searchr1,song2025r1searcher}. Successful trajectories provide limited guidance for context-induced deviations, while outcome rewards offer coarse credit over long interactions. To provide finer-grained supervision, recent methods introduce intermediate signals for query quality, information acquisition, or failure-relevant actions~\citep{wang2025stepsearch,ma2026sd,wen2026smartsearch,yeo2026hint}. Yet an action may appear locally appropriate while continuing a trajectory shaped by misleading context, allowing the search to remain coherent while progressively departing from the original search target.

We propose \textbf{BOUND}, a brief-guided corrective preference distillation framework for persistent search drift. Starting from student-generated rollouts, BOUND constructs a search-state brief that records the original search target, key constraints, confirmed evidence, missing information, and drift status. The brief serves as a teacher-side privileged representation during preference construction. By preserving the original search target and key constraints, the brief provides a stable task-level reference when search context may reinforce a misleading entity, a dropped constraint, or a displaced topic.

Guided by the brief, the teacher evaluates each decision-time state and the student's original continuation against the task and currently available evidence, determining whether a reliable local preference can be established. This assessment, together with the rollout outcome, determines which contrast to construct. For an erroneous continuation, BOUND contrasts a student-specific correction with the original continuation; for a supported answer, it contrasts the answer with an unnecessary retrieval alternative. States without a reliable preference contribute no contrast. Because both continuations share the same student-visible state, the preference isolates the local search-control decision, exposing how a locally plausible continuation may reinforce drift while the preferred alternative remains anchored to the task and evidence. Each contrast is represented as a state-matched preference pair, and each validated pair operationalizes a search-control boundary where the alternatives imply different local decisions. We use direct preference optimization (DPO)~\citep{rafailov2023direct} to distill these preferences into the student, enabling it to correct emerging search drift. The brief and teacher-side computation remain confined to training.

We evaluate BOUND on multi-hop question answering and deep search benchmarks. BOUND consistently outperforms Trajectory SFT under the same Qwen3-4B-Instruct-2507 initialization, as well as the comparably sized ORBIT-4B baseline. Compared with Trajectory SFT, BOUND achieves gains of 5.6 EM points on Bamboogle and 4.8 accuracy points on BrowseComp-Plus. Analyses demonstrate the value of anchoring teacher assessment to the original search target and key constraints, combining brief-guided assessment with rollout outcomes during preference construction, and constructing corrections that target the student's original continuation.

Our main contributions are summarized as follows:
\begin{itemize}
\item We formulate persistent search drift as a cumulative search-control failure where locally plausible evidence redirects decisions from the original search target or constraints. We characterize three representative forms of this failure: wrong-anchor drift, constraint drift, and local-topic drift.

\item We propose BOUND, which combines brief-guided assessment with rollout outcomes to construct reliable local contrasts and distill student-specific preferences. Each validated state-matched pair operationalizes a search-control boundary.

\item We demonstrate consistent gains across multi-hop question answering and deep search benchmarks, with analyses showing the value of brief-guided assessment, its joint use with rollout outcomes in preference construction, and student-specific corrective preferences.
\end{itemize}

\section{Related Work}

\subsection{LLM-based Deep Search}

LLM-based deep search interleaves querying, retrieval, and reasoning for multi-step information seeking. Early work established this reasoning-and-search paradigm~\citep{nakano2021webgpt,yao2023react,trivedi2023interleaving}, while later agentic retrieval frameworks extended it to more complex search settings~\citep{li2025searcho1}. These advances strengthen overall search capability, but repeated retrieval and reasoning can accumulate partially relevant evidence and gradually divert the search from the original information need, making search drift a recurring challenge.

\begin{figure*}[t]
    \centering
    \includegraphics[width=1\linewidth]{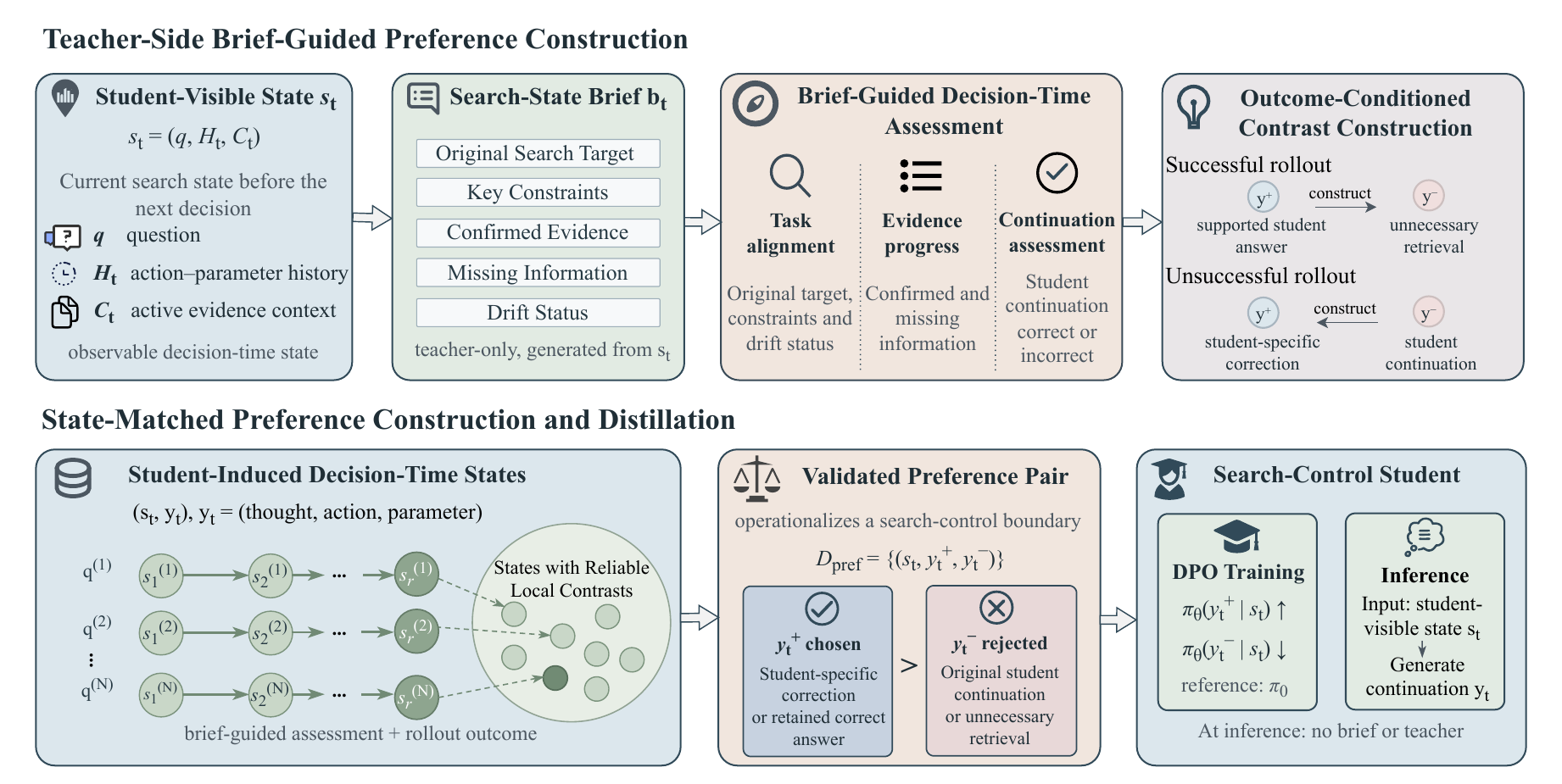}
    \caption{Overview of BOUND. A teacher-side brief and rollout outcomes yield validated state-matched preference pairs that operationalize search-control boundaries and are distilled into the student.}
    \Description{BOUND combines brief-guided assessment with rollout outcomes to construct validated state-matched pairs and distills them into the student without teacher-side information at inference.}
    \label{fig:method}
\end{figure*}

\subsection{Training Methods for Search Control}

Search control can be learned from supervised trajectories constructed for challenging or shortcut-resistant multi-step tasks~\citep{li2025websailor,deng2026fort}, through outcome-based reinforcement learning~\citep{jin2025searchr1,song2025r1searcher}, or by combining distillation with policy optimization~\citep{kotoge2026dgpo}. More localized supervision comes from preference learning over alternative behaviors~\citep{ouyang2022training,bai2022training,rafailov2023direct}, including retrieval-feedback-driven preferences for query expansion~\citep{li2026retrieval}. Recent methods further introduce query-level process rewards, failure-relevant action selection, stepwise objectives, first-error correction, environment feedback, grouped-rollout hindsight, action-centered rescoring, or teacher--student discrepancies~\citep{wen2026smartsearch,yeo2026hint,wang2025stepsearch,li2026and,ma2026sd,zhang2026stepopsd,zhong2026sod}. DAS targets over- and under-search through search-versus-answer preferences~\citep{zhang2026search}, while CSO constructs shared-state preference pairs from failed rollouts using outcome-changing alternatives~\citep{li2026verified}. However, retrieval-side drift remains underexplored: locally plausible continuations can steer search away from the original target, key constraints, or unresolved evidence needs.

\subsection{Teacher-Guided Distillation with Privileged Information}

LLM distillation transfers reasoning traces from stronger teachers to smaller models~\citep{ho2023large,hsieh2023distilling,shridhar2023distilling}. PLaD instead constructs pseudo-preference pairs from teacher and student outputs~\citep{zhang24plad}. On-policy training and self-distillation further supervise student trajectories to reduce the mismatch between training and inference~\citep{zhao2026selfdistilled,shenfeld2026selfdistillation}. $\pi$-Distill jointly trains a shared-parameter PI-conditioned teacher and unconditioned student, while OPSD transfers teacher behavior through reverse-KL-regularized reinforcement learning for multi-turn agents~\citep{penaloza2026privileged}. Skill-SD exposes trajectory-derived skills only to the teacher during distillation~\citep{wang2026skill}. These methods show that teacher-only information can provide richer supervision without increasing inference-time inputs.

\section{Method}
\label{sec:method}

\subsection{Overview}
\label{sec:method-overview}

BOUND addresses persistent search drift by distilling corrective preferences at search-control boundaries. Such boundaries arise when alternative continuations from the same student-visible state have different search-control implications and decision-time evidence supports a reliable preference between them. BOUND operationalizes each boundary through a validated state-matched pair, avoiding the assumption that every decision in an unsuccessful rollout is necessarily erroneous.

Figure~\ref{fig:method} summarizes the workflow. For each state, the teacher produces a search-state brief covering the original search target, key constraints, confirmed evidence, missing information, and drift status. The brief anchors decision-time assessment around task alignment, evidence progress, and the student's original continuation. Guided by the brief, the teacher determines whether the continuation contains a local search-control error with a preferable correction. Together with the rollout outcome, this assessment determines pair form: an unsuccessful rollout yields the erroneous continuation and its student-specific correction, whereas a successful one yields the earliest evidence-supported \textsc{Answer} and an unnecessary retrieval continuation. States without a sufficiently reliable preference yield no pair.

Each validated pair shares the same student-visible state but contrasts continuations expressing different local search-control decisions, with the chosen continuation reliably preferable under the task and available evidence at that decision point. Using the initial policy as reference, DPO uses these preferences to update the student toward task- and evidence-aligned behavior and prevent local deviations from compounding across later search steps. The brief and teacher-side computation remain confined to training; inference requires only the student-visible state.

\subsection{Brief-Guided Corrective Preference Distillation}
\label{sec:corrective-distillation}

\subsubsection{Corrective Preferences over Student-Induced States}
\label{sec:process-distillation}

Given a training question $q$, the initial student policy $\pi_0$ interacts with the retrieval environment to generate a trajectory
\[
\tau^{\pi_0}(q)
=
\left[(s_t,y_t^0,o_t)\right]_{t=1}^{T}.
\]
At step~$t$, the student-visible state is $s_t=(q,H_t,C_t)$, where $H_t$ denotes the preceding actions and their parameters, $C_t$ is the active evidence context, and $y_t^0$ is the student's original continuation. If $y_t^0$ invokes retrieval, $o_t$ is the resulting observation; otherwise, $o_t=\varnothing$. Here, $T$ denotes the number of decision steps in the trajectory.

Persistent drift emerges and propagates through these student-induced states. An unsupported entity may become the anchor of subsequent queries, a key constraint may disappear during reformulation, or a locally coherent side topic may displace the unresolved information need. Later continuations can therefore remain locally plausible while progressively departing from the original search target and key constraints.

BOUND performs correction over states induced by the student's own policy rather than requiring the student to imitate complete teacher trajectories. These states expose errors and ambiguities arising during the student's actual search process, including situations absent from an ideal teacher trajectory. For each state $s_t$ admitting a reliable local contrast, BOUND defines a local preference
\[
y_t^+ \succ_{s_t} y_t^-,
\]
where $y_t^+$ is preferred to $y_t^-$ under the same state $s_t$. The two continuations share the same question, preceding interaction history, and active evidence context. Their comparison therefore isolates the preferred local decision rather than differences in available information. This formulation transfers a state-specific local preference without requiring the teacher to regenerate the remaining trajectory.

\subsubsection{Search-State Briefs as Privileged Information}
\label{sec:brief-guide}

Reliable correction requires assessing the student-visible state relative to the original search target and key constraints. However, the accumulated context may already contain misleading evidence or assumptions that contribute to drift, causing the teacher to reproduce the same local bias.

For each decision-time state, BOUND constructs a structured search-state brief
\[
b_t
=
g_{\mathrm{B}}(s_t)
=
g_{\mathrm{B}}(q,H_t,C_t),
\]
as a teacher-side privileged, task-anchored state representation. It contains five fields: \textbf{Original Search Target}, \textbf{Key Constraints}, \textbf{Confirmed Evidence}, \textbf{Missing Information}, and \textbf{Drift Status}. The first two preserve a stable reference to the original task, while the remaining fields summarize evidence progress and whether the current search remains aligned with that reference.

By separating task-invariant requirements from state-dependent search progress, the brief makes explicit what should remain fixed and what still needs to be resolved. It therefore provides a task-anchored reference for decision-time assessment, allowing the teacher to evaluate the student's continuation against the original search objective rather than relying solely on the potentially
drifted trajectory context. In this way, the brief helps distinguish continuations that are merely locally plausible from those that remain aligned with the unresolved information need and key constraints.

The brief and the associated decision-time assessment are used only for
preference construction during training. BOUND then optimizes the student over the resulting preference dataset using DPO, with the initial policy $\pi_0$ as the reference policy. Because the student itself remains conditioned only on $s_t$, the learned policy requires neither the brief nor any teacher-side computation during inference.

\subsection{Corrective Preference Construction at Search-Control Boundaries}
\label{sec:preference-construction}

BOUND constructs corrective preferences in two stages. It first combines brief-guided decision-time assessment with rollout outcomes to determine the contrast for each student-induced state (Section~\ref{sec:decision-assessment}). In unsuccessful rollouts, an erroneous continuation is paired with a student-specific correction. In successful rollouts, the earliest supported answer is paired with an unnecessary retrieval alternative. Each retained contrast forms a state-matched preference pair whose continuations encode different local search decisions under the same student-visible state (Section~\ref{sec:local-preference}), thereby operationalizing a search-control boundary.

\begin{algorithm}[t]
\caption{Corrective preference construction}
\label{alg:preference-construction}
\begin{algorithmic}[1]
\Require Rollout $\tau=[(s_t,y_t^0,o_t)]_{t=1}^{T}$ and outcome $r\in\{0,1\}$, where $r=1$ denotes a correct final answer
\Ensure Preference dataset $\mathcal{D}_{\mathrm{pref}}$
\State $\mathcal{D}_{\mathrm{pref}}\gets\emptyset$
\For{$t=1,\ldots,T$}
\State $b_t\gets g_{\mathrm{B}}(s_t)$
\State $(d_t,\widetilde{y}_t^{\mathrm{T}})
\gets\operatorname{AssessAndCorrect}(s_t,b_t,y_t^0)$
\Comment{$d_t=(z_t,\rho_t,e_t)$, possibly with $\widetilde{y}_t^{\mathrm{T}}=\varnothing$}
\If{$r=0 \land e_t
\land z_t\neq\texttt{not\_selected}
\land \widetilde{y}_t^{\mathrm{T}}\neq\varnothing$}
\State $y_t^+\gets\widetilde{y}_t^{\mathrm{T}}$
\State $y_t^-\gets y_t^0$
\If{the pair passes validation}
\State Add $(s_t,y_t^+,y_t^-)$ to $\mathcal{D}_{\mathrm{pref}}$
\EndIf
\ElsIf{$r=1 \land \operatorname{SupportedAnswer}(d_t,y_t^0)$}
\State $y_t^+\gets y_t^0$
\State $y_t^-\gets\operatorname{UnnecessaryRetrieval}(s_t,b_t,y_t^0)$
\If{the pair passes validation}
\State Add $(s_t,y_t^+,y_t^-)$ to $\mathcal{D}_{\mathrm{pref}}$
\EndIf
\State \textbf{break}
\EndIf
\EndFor
\State \Return $\mathcal{D}_{\mathrm{pref}}$
\end{algorithmic}
\end{algorithm}

Before entering the preference dataset, each pair is checked for valid format, consistency with its construction route, a difference in action or parameter, and the absence of teacher-only information. For supported-answer pairs, the rejected continuation must invoke further retrieval. Pairs that pass these checks are retained for preference optimization.

Algorithm~\ref{alg:preference-construction} summarizes the construction process. Lines~3 to~4 build the brief and assess the original continuation. Lines~5 to~10 form corrective pairs for unsuccessful rollouts, while Lines~11 to~17 form a termination pair from the earliest supported answer in a successful rollout. The auxiliary tag $z_t$ serves as a teacher-side routing signal, while pair retention follows the construction route, rollout outcome, local assessment, correction availability, and validation checks. Validation removes malformed pairs, pairs that differ only in reasoning text, pairs inconsistent with their construction route, and pairs that expose teacher-only information. Appendix~\ref{app:preference-construction} details the interfaces, schemas, generation requirements, and validation checks. Exact templates appear in the released code.

\subsubsection{Decision-Time Assessment with Rollout Outcomes}
\label{sec:decision-assessment}

BOUND combines brief-guided decision-time assessment with rollout outcomes to determine whether a reliable local pair can be constructed and which contrast form should be used. Not every state in an unsuccessful rollout provides useful supervision. A continuation may be locally appropriate even when the final answer is incorrect, since other states may still require additional evidence, an evidence-supported answer, or recovery of a displaced target or constraint. BOUND therefore assesses each state independently rather than propagating the rollout outcome across the trajectory.

For each state, the teacher examines the student-visible state, search-state brief, and original continuation to determine whether the continuation contains a specific local search-control error with a clearly preferable correction. The observation produced by the continuation is excluded, so the assessment relies only on information available at decision time rather than later trajectory information. If no specific error with a reliable correction can be established, the teacher abstains and constructs no corrective pair.

The rollout outcome determines the permitted contrast form rather than the local error judgment. For an unsuccessful rollout, BOUND pairs an erroneous continuation with its student-specific correction, while locally appropriate continuations produce no pair. For a successful rollout, it retains the earliest evidence-supported correct \textsc{Answer} and contrasts it with an unnecessary retrieval continuation. In both cases, the two continuations share the same decision-time state, and the resulting pair must pass validation before entering preference optimization.

\subsubsection{State-Matched Preference Construction}
\label{sec:local-preference}

Each eligible state yields a state-matched preference pair
\[
\left(x_t,y_t^+,y_t^-\right),
\qquad
x_t=s_t.
\]
Both continuations share the same question, action history, and active evidence context, so their comparison isolates the preferred local decision.

When the student's original continuation is judged incorrect, the teacher generates a student-specific correction conditioned on the state, brief, and continuation:
\[
(y_t^+,y_t^-)
=
(\widetilde{y}_t^{\mathrm{T}},y_t^0).
\]
The correction may seek unresolved evidence, restore a displaced search target or missing constraint, or answer when the evidence is sufficient. Conditioning on $y_t^0$ targets the observed error rather than producing a generic teacher continuation.

For the earliest supported correct answer in a successful rollout, BOUND constructs
\[
(y_t^+,y_t^-)
=
\left(
y_t^0,
\operatorname{UnnecessaryRetrieval}(s_t,b_t,y_t^0)
\right),
\]
where the rejected continuation is plausible yet unnecessary. This preserves evidence-supported termination and discourages further search once the evidence establishes the answer.

During execution, selecting \textsc{Reroute} implements the corrected search decision by appending the action and revised query to the history and advancing the student-visible state through a fixed environment transition. The retrieved passages are added while preserving prior history and evidence. The same transition is used during rollout collection and inference, so rerouting supervision matches the action actually executed, as detailed in Appendix~\ref{app:preference-serialization}.

Each continuation contains a thought, action, and parameter, but only the action and parameter enter history. BOUND removes malformed pairs, pairs lacking distinct local search-control decisions, and outputs exposing the brief, assessment record, rollout outcome, or other teacher-only information. For a supported correct answer, the rejected continuation must invoke retrieval rather than provide another answer. The surviving pairs form
\[
\mathcal{D}_{\mathrm{pref}}
=
\left\{
\left(x_i,y_i^+,y_i^-\right)
\right\}_{i=1}^{N},
\]
where $N$ is the number of validated preference pairs.

\section{Experiments} 
\label{sec:experiments}

\begin{table*}[t]
\centering
\footnotesize
\caption{Main results on multi-hop and deep search benchmarks. All values are percentages.}
\label{tab:main-results}

\scalebox{1.227}{
\begin{tabular}{lcccccccccccccc}
\toprule
\multirow{2.5}{*}{\textbf{Method}}
& \multicolumn{2}{c}{HotpotQA}
& \multicolumn{2}{c}{MuSiQue}
& \multicolumn{2}{c}{2Wiki}
& \multicolumn{2}{c}{Bamboogle}
& \multicolumn{3}{c}{FRAMES}
& \multicolumn{3}{c}{GAIA} \\
\cmidrule(lr){2-3}
\cmidrule(lr){4-5}
\cmidrule(lr){6-7}
\cmidrule(lr){8-9}
\cmidrule(lr){10-12}
\cmidrule(lr){13-15}
& \textbf{EM} & \textbf{F1}
& \textbf{EM} & \textbf{F1}
& \textbf{EM} & \textbf{F1}
& \textbf{EM} & \textbf{F1}
& \textbf{EM} & \textbf{CEM} & \textbf{F1}
& \textbf{EM} & \textbf{CEM} & \textbf{F1} \\
\midrule
\multicolumn{15}{l}{\textit{Qwen3-4B-Instruct without task-specific search training}} \\
Direct Inference
& 18.4 & 26.8
& 2.0 & 7.4
& 18.1 & 25.1
& 5.6 & 14.1
& 2.3 & 4.9 & 8.1
& 2.9 & 3.9 & 5.8 \\

Standard RAG
& 24.0 & 32.9
& 4.3 & 10.1
& 20.2 & 29.8
& 8.8 & 17.7
& 4.0 & 6.4 & 11.2 
& -- & -- & -- \\

IRCoT
& 26.1 & 39.9
& 5.1 & 10.9
& 20.8 & 29.7
& 24.0 & 35.9
& 4.4 & 8.6 & 11.7
& 14.6 & 19.4 & 28.6 \\
ReAct
& 29.6 & 41.7
& 7.9 & 14.2
& 25.8 & 34.6
& 28.0 & 37.2
& 8.4 & 9.2 & 15.2
& 16.5 & 17.5 & 22.6 \\
\midrule
\multicolumn{15}{l}{\textit{Trained search baselines based on Qwen2.5-7B models}} \\
Search-R1
& 37.4 & 49.8
& 12.2 & 19.9
& 33.7 & 40.3
& 32.0 & 42.0
& 12.4 & 13.5 & 19.5
& 17.5 & 19.4 & 30.6 \\
StepSearch
& 36.0 & 46.1
& 13.0 & 23.7
& 39.6 & 46.6
& 32.0 & 42.9
& 12.5 & 13.2 & 19.1
& 17.5 & 20.4 & 29.9 \\
SimpleDeepSearcher
& 39.6 & 51.9
& 12.6 & 23.1
& 48.1 & 55.5
& 34.4 & 47.0
& 13.2 & 13.7 & 19.9
& 24.3 & 25.2 & 34.3 \\
R1-Searcher
& 41.1 & 53.5
& 15.1 & 24.7
& \textbf{48.9} & \textbf{55.8}
& 37.6 & 46.5
& 12.0 & 12.9 & 19.3
& 16.5 & 18.5 & 28.3 \\
\midrule
\multicolumn{15}{l}{\textit{Trained search methods based on Qwen3-4B models}} \\
ORBIT-4B
& 35.7 & 45.5
& 12.5 & 20.2
& 38.6 & 45.4
& 39.2 & 46.7
& 12.5 & 15.4 & 20.4
& 19.4 & 24.3 & 28.0 \\
\textbf{BOUND}
& \textbf{42.4} & \textbf{54.0}
& \textbf{15.9} & \textbf{26.3}
& 47.7 & 53.8
& \textbf{42.4} & \textbf{54.2}
& \textbf{14.2} & \textbf{15.8} & \textbf{24.7}
& \textbf{27.2} & \textbf{30.1} & \textbf{40.8} \\
\bottomrule
\end{tabular}
}
\end{table*}

\subsection{Experimental Settings}

\paratitle{Datasets.}
We derive training preferences from existing and auxiliary synthetic multi-hop questions. Existing questions are sampled and screened from the official HotpotQA~\citep{yang2018hotpotqa} and MuSiQue~\citep{trivedi2022musique} training splits, while synthetic questions are constructed from multi-path evidence in a fixed English Wikipedia dump. No evaluation questions, answers, annotations, or benchmark-specific supervision are used. Appendix~\ref{app:synthetic-data} details the construction and contamination audit.

We evaluate on the development sets of HotpotQA, MuSiQue, and 2WikiMultiHopQA~\citep{ho2020constructing}, the full Bamboogle benchmark~\citep{press2023measuring}, and FRAMES~\citep{krishna2025fact}, GAIA~\citep{mialon2024gaia}, and BrowseComp-Plus~\citep{chen2025browsecomp}. Following SimpleDeepSearcher~\citep{sun2025simpledeepsearcher}, GAIA uses its 103-example text-only validation subset.

\paratitle{Baselines.}
We compare BOUND against prompting-based and trained search agents. Direct Inference uses no retrieval, whereas Standard RAG~\citep{lewis2020retrieval} performs a single retrieval step. ReAct~\citep{yao2023react} alternates reasoning with search, while IRCoT~\citep{trivedi2023interleaving} interleaves chain-of-thought reasoning with retrieval. For trained 7B baselines, we evaluate the Qwen2.5-7B variants of Search-R1~\citep{jin2025searchr1}, StepSearch~\citep{wang2025stepsearch}, SimpleDeepSearcher~\citep{sun2025simpledeepsearcher}, and R1-Searcher~\citep{song2025r1searcher}. We additionally evaluate ORBIT-4B~\citep{thakur2026orbit} as a trained 4B baseline. Together, these baselines cover no-retrieval inference, single-step retrieval, iterative search prompting, and trained multi-step search.

We rerun the baselines in Table~\ref{tab:main-results} under our experimental setup, following their original formulations and released implementations where available. All methods use the same retrieval environment and top-5 retrieval setting. On the multi-hop benchmarks, FRAMES, and GAIA, iterative-search methods are limited to 10 steps per question and use a generation temperature of 0.6. BrowseComp-Plus is evaluated separately with its fixed corpus and official evaluator. The external reference systems reported in Table~\ref{tab:bcplus-results} are taken directly from the public leaderboard.

\paratitle{Evaluation Metrics.}
For multi-hop QA benchmarks, we report exact match (EM) and token-level F1 following common practice. For FRAMES and GAIA, we additionally report cover exact match (CEM), which counts a prediction as correct if any normalized gold answer or accepted alias appears in the normalized final response, allowing minor variations in answer phrasing.

For BrowseComp-Plus, we use its fixed corpus and official leaderboard evaluator, reporting answer accuracy judged by Qwen3-32B together with retrieval recall under the standard metric definitions.

\paratitle{Implementation Details.}
Qwen3-4B-Instruct-2507~\citep{yang2025qwen3} serves as the student, while DeepSeek-V4-Flash~\citep{deepseekai2026deepseekv4} conducts brief-guided assessment and preference construction. We refer to the student as Qwen3-4B-Instruct throughout the paper. Training uses 2,424 step-level preference pairs constructed from decision-time states in student rollouts. Appendix~\ref{app:synthetic-data} details the question sources and synthetic construction procedure. We train with DPO for one epoch using a learning rate of $1\mathrm{e}{-6}$, $\beta=0.1$, and a maximum sequence length of 8,192 tokens, with the initial checkpoint serving as the reference policy. The search-state brief is used only for preference construction and is unavailable at inference time.

For HotpotQA, MuSiQue, 2WikiMultiHopQA, Bamboogle, and FRAMES, we retrieve over the March 1, 2022 English Wikipedia dump using BM25~\citep{robertson2009probabilistic} and E5-base-v2~\citep{wang2022text} with reciprocal rank fusion~\citep{cormack2009reciprocal}. For BrowseComp-Plus, we use its fixed corpus with Qwen3-Embedding-8B~\citep{qwen3embedding}. Following prior work~\citep{sun2025simpledeepsearcher}, GAIA uses a shared Serper API and frozen QwQ-32B summarizer.

\subsection{Main Results}
\label{sec:main-results}

\paratitle{Overall Performance.}
Table~\ref{tab:main-results} reports the main results on multi-hop QA and deep search benchmarks. Initialized from Qwen3-4B-Instruct and evaluated under the shared retrieval and inference settings, BOUND achieves the best performance on five of the six datasets. We make three observations.

(1) \textbf{Inference-time retrieval improves performance but remains limited.}
Standard RAG, IRCoT, and ReAct generally outperform Direct Inference, demonstrating the value of external evidence and iterative retrieval. However, they remain behind methods explicitly trained for multi-step search. BOUND outperforms both IRCoT and ReAct on every reported metric, with substantial gains across the six datasets, including Bamboogle and GAIA. These results suggest that retrieval alone does not fully address the need to control evidence acquisition and stopping.

(2) \textbf{Training for search control yields clear improvements.}
Trained search methods generally outperform inference-time retrieval baselines, highlighting the value of task-specific optimization for multi-step search. Among the trained methods, BOUND achieves the best performance on five of the six datasets. It outperforms ORBIT-4B, a Qwen3-4B search model trained with GRPO, on every reported metric and surpasses the trained Qwen2.5-7B baselines on five datasets. These comparisons support the effectiveness of local preference supervision for search control.

(3) \textbf{BOUND performs strongly on challenging deep search benchmarks.}
BOUND achieves 24.7 F1 on FRAMES and 40.8 F1 on GAIA, exceeding the strongest 7B baseline by 4.8 and 6.5 F1 points, respectively, under the shared evaluation setting. These results are consistent with BOUND's focus on maintaining the target and constraints, assessing unresolved evidence needs, and stopping when sufficient evidence has been acquired.

The main exception is 2Wiki, where SimpleDeepSearcher and R1-Searcher achieve higher EM and F1 scores. Unlike BOUND, both methods incorporate 2Wiki questions into their training-data construction~\citep{sun2025simpledeepsearcher,song2025r1searcher}. Nevertheless, despite its smaller 4B scale and lack of 2Wiki training questions, BOUND remains competitive with the strongest 7B baselines and outperforms ORBIT-4B by 9.1 EM and 8.4 F1 points. This result indicates transfer beyond the datasets used as sources of its training questions.

\begin{table}[t]
\centering
\small
\caption{Results on BrowseComp-Plus. Public leaderboard systems using Qwen3-Embedding-8B are included for reference.}
\label{tab:bcplus-results}
\scalebox{1.06}{
\begin{tabular}{lccc}
\toprule
\textbf{Method} & \textbf{Size} & \textbf{Acc. (\%)} & \textbf{Recall} (\%) \\
\midrule
Claude Sonnet 4 & -- & 37.4 & 47.3 \\
Claude Opus 4 & -- & 36.8 & 50.8 \\
Gemini-2.5-Flash & -- & 34.6 & 40.2 \\
Gemini-2.5-Pro & -- & 29.5 & 35.3 \\
\midrule
kimi-k2-0711-preview & 1000B & 35.4 & 38.4 \\
DeepSeek-R1-0528 & 685B & 16.4 & 16.3 \\
Search-R1-32B & 32B & 11.1 & 10.2 \\
Tongyi-DeepResearch-30B-A3B & 30B & 44.5 & 62.3 \\
WebSailor-32B & 32B & 27.6 & 29.3 \\
oss-20B-high & 20B & 35.1 & 49.3 \\
FRUGALRAG-7B & 7B & 20.5 & 23.6 \\
\midrule
\textbf{BOUND} & 4B & 29.6 & 32.1 \\
\bottomrule
\end{tabular}}
\end{table}

\paratitle{Performance on BrowseComp-Plus.}
Table~\ref{tab:bcplus-results} reports results on BrowseComp-Plus, a challenging deep search benchmark with a fixed search corpus. The selected public reference systems use Qwen3-Embedding-8B and return the top five documents at each search step. BOUND uses the same retrieval configuration and official evaluator. Trained only on multi-hop and synthetic search questions, BOUND achieves 29.6 answer accuracy under the Qwen3-32B judge and 32.1 retrieval recall. It exceeds DeepSeek-R1-0528, Search-R1-32B, WebSailor-32B, and FRUGALRAG-7B on both metrics. Although BOUND remains below the strongest proprietary models and large-scale deep-research systems, these results indicate transfer to harder search questions despite its compact scale and without using BrowseComp-Plus training data. The teacher model and search-state brief are not required at inference time. Additional experiments with Qwen3.5-4B in Appendix~\ref{app:qwen35-results} show consistent improvements in both metrics.

\subsection{Ablation Study}
\label{sec:ablation}

Table~\ref{tab:ablation} evaluates the main training components and execution choices of BOUND. All variants share the same training questions, candidate state pool, student backbone, and optimization setup. The remaining supervision-construction procedure is kept unchanged, so variations in the resulting preference sets arise directly from the ablated component rather than from resampling or independently constructed training data. In \emph{w/o Rerouting Supervision}, the teacher is restricted to \textsc{Continue} and \textsc{Answer}. In \emph{w/o Teacher Correction}, teacher corrections are replaced with student-generated alternatives, while the states, briefs, assessments, and rejected continuations remain unchanged. In \emph{w/o Student-Specific Correction}, corrections are generated without access to the student's original continuation or the identified local error. \emph{Query-Only Rerouting} uses the same preference pairs as Full BOUND and disables only the filtering of passages from the immediately preceding retrieval during execution.

On BrowseComp-Plus, removing rerouting supervision produces the largest accuracy drop: judge accuracy falls from 29.6 to 21.2, while retrieval recall falls from 32.1 to 26.4. This indicates that explicit re-anchoring is important when early deviations persist across subsequent search steps.

Replacing teacher corrections with student-generated alternatives reduces BrowseComp-Plus judge accuracy to 21.7 and retrieval recall to 25.8. The result suggests that teacher corrections more reliably preserve the original target and key constraints while directing the student toward missing information.

Without student-specific correction, Bamboogle F1 drops from 54.2 to 52.3, and BrowseComp-Plus judge accuracy falls from 29.6 to 24.6. Conditioning the correction on the student's continuation and the identified local error is therefore more effective than generating a generic continuation from the state alone.

Query-only rerouting reduces BrowseComp-Plus judge accuracy from 29.6 to 26.0 and retrieval recall from 32.1 to 27.7. This suggests that removing passages associated with the abandoned search direction helps the rerouting query guide subsequent evidence acquisition.

\begin{table}[t]
    \centering
    \small
    \caption{Ablation results on Bamboogle and BrowseComp-Plus.}
    \label{tab:ablation}
    \scalebox{1.02}{
    \begin{tabular}{lcccc}
        \toprule
        \multirow{2.5}{*}{\textbf{Variant}}
        & \multicolumn{2}{c}{Bamboogle}
        & \multicolumn{2}{c}{BrowseComp-Plus} \\
        \cmidrule(lr){2-3}
        \cmidrule(lr){4-5}
        & \textbf{EM} & \textbf{F1} & \textbf{Acc.} & \textbf{Recall} \\
        \midrule
        w/o Rerouting Supervision
        & 37.6 & 50.3 & 21.2 & 26.4 \\
        w/o Teacher Correction
        & 36.8 & 50.8 & 21.7 & 25.8 \\
        w/o Student-Specific Correction
        & 38.4 & 52.3 & 24.6 & 28.6 \\
        Query-Only Rerouting
        & 38.4 & 51.7 & 26.0 & 27.7 \\
        \midrule
        \textbf{BOUND}
        & \textbf{42.4}
        & \textbf{54.2}
        & \textbf{29.6}
        & \textbf{32.1} \\
        \bottomrule
    \end{tabular}}
\end{table}

\subsection{Comparison with Trajectory-Level Distillation}
\label{sec:trajectory-distillation}
We compare BOUND with Trajectory supervised fine-tuning (Trajectory SFT), which imitates teacher-generated search trajectories with token-level cross-entropy. Both methods share the training questions, teacher, student initialization, action space, retriever, and evaluation protocol throughout the controlled comparison. Their initial policy is the same Qwen3-4B-Instruct checkpoint evaluated under our search-control interface and environment transitions, which differs from the prompting-based setup used in Table~\ref{tab:main-results}. This comparison isolates complete-trajectory SFT from state-matched DPO supervision.

\begin{table}[t]
\centering
\small
\caption{Comparison with trajectory-level distillation under our search-control interface.}
\label{tab:trajectory-distillation}
\scalebox{1.12}{
\begin{tabular}{lcccc}
\toprule
\multirow{2.5}{*}{\textbf{Method}}& \multicolumn{2}{c}{Bamboogle}
& \multicolumn{2}{c}{BrowseComp-Plus} \\
\cmidrule(lr){2-3}
\cmidrule(lr){4-5}
 & \textbf{EM} & \textbf{F1} & \textbf{Acc.} & \textbf{Recall} \\
\midrule
Initial Search Policy
& 28.8 & 36.5 & 14.9 & 14.0 \\
Trajectory SFT
& 36.8 & 47.7 & 24.8 & 30.4 \\
BOUND
& \textbf{42.4} & \textbf{54.2}
& \textbf{29.6} & \textbf{32.1} \\
\bottomrule
\end{tabular}}
\end{table}
Table~\ref{tab:trajectory-distillation} shows that trajectory imitation substantially improves the initial search policy, while BOUND further gains 5.6 EM and 6.5 F1 on Bamboogle and 4.8 accuracy and 1.7 recall on BrowseComp-Plus. The BrowseComp-Plus accuracy gain is statistically significant ($p<0.001$). The greater improvement in accuracy relative to recall suggests that BOUND benefits not only from acquiring relevant evidence but also from making better search-control and answering decisions over the available evidence.

\begin{table*}[t]
\centering
\footnotesize
\renewcommand{\arraystretch}{1.15}
\caption{Representative cases of persistent search drift.}
\label{tab:trajectory-case-study}
\begin{tabularx}{\textwidth}{
    >{\raggedright\arraybackslash}p{0.13\textwidth}
    >{\raggedright\arraybackslash}p{0.21\textwidth}
    >{\raggedright\arraybackslash}X
    >{\raggedright\arraybackslash}X
}
\toprule
\textbf{Drift Type}
& \textbf{Task Cue}
& \textbf{Trajectory SFT}
& \textbf{BOUND} \\
\midrule
\textbf{Wrong-Anchor Drift}
&
Identify the surviving spouse who founded a philanthropic foundation.
&
Introduces an unsupported celebrity association and makes
\emph{Jeff Bezos} the search anchor.

\textbf{Answer:} Jeff Bezos
{\color{red!70!black}\ding{55}}
&
Preserves the original target and verifies the foundation and role
constraints.

\textbf{Answer:} Trudy Bronner
{\color{green!50!black}\ding{51}}
\\
\midrule
\textbf{Constraint Drift}
&
Identify a game released during 2001--2007 under the given developer
and publisher constraints.
&
Drops the release-window constraint during query reformulation.

\textbf{Answer:} It Takes Two
{\color{red!70!black}\ding{55}}
&
Retains the release window and the developer and publisher constraints
throughout the search.

\textbf{Answer:} Law \& Order: Justice Is Served
{\color{green!50!black}\ding{51}}
\\
\midrule
\textbf{Local-Topic Drift}
&
Identify the insect associated with the dormant volcano specified by
the satellite and caldera clues.
&
Follows a coherent but displaced ``stick insect'' side path.

\textbf{Answer:} Lord Howe stick insect
{\color{red!70!black}\ding{55}}
&
Returns to the unresolved volcano evidence and redirects search away
from the displaced topic.

\textbf{Answer:} Mosquito
{\color{green!50!black}\ding{51}}
\\
\bottomrule
\end{tabularx}
\end{table*}

\paratitle{Case Study.}
Table~\ref{tab:trajectory-case-study} presents representative cases of divergence between Trajectory SFT and BOUND for the three drift manifestations introduced in Section~\ref{sec:introduction}.

Across the three cases, Trajectory SFT follows locally plausible continuations after the search state becomes misaligned, allowing an unsupported entity, a dropped constraint, or a side topic to redirect later search. BOUND instead preserves or restores the target, constraints, or unresolved evidence need while recovery remains possible, before these local errors become entrenched in the trajectory. Its state-matched preferences contrast the drift-inducing decision with a corrective alternative that re-anchors the target, reinstates the constraint, or redirects search toward missing evidence. Together with the controlled ablations, these comparisons support the interpretation that BOUND improves performance by correcting the local search-control decisions through which drift propagates across subsequent steps.

\subsection{Effect of Search-State Briefs on Teacher Assessment and Correction}
\label{sec:teacher-brief-analysis}

\paratitle{Brief quality.}
Manual inspection of 50 randomly sampled search-state briefs shows reliable preservation of the original target and key constraints, with most errors occurring in missing-information summaries and drift-status assessment. Field-level results and evaluation criteria appear in Appendix~\ref{app:brief-quality}.

\paratitle{Student-level effect.}
We next examine whether search-state briefs improve the downstream utility of teacher-generated preference data. Using the same student-induced decision-time states, we build preference pairs with DeepSeek-V4-Flash and Qwen3-32B, each with and without briefs. All resulting Qwen3-4B students are trained and evaluated under the same configuration.

\begin{table}[t]
\centering
\small
\caption{Effect of search-state briefs on student performance under different teachers. DeepSeek refers to DeepSeek-V4-Flash, and $\Delta$ denotes the absolute gain from adding the brief.}
\begin{tabular}{lccccc}
\toprule
\multirow{2.5}{*}{\textbf{Teacher}} & \multirow{2.5}{*}{\textbf{Brief}}
& \multicolumn{2}{c}{BrowseComp-Plus}
& \multicolumn{2}{c}{Bamboogle} \\
\cmidrule(lr){3-4}\cmidrule(lr){5-6}
& & \textbf{Acc.} & \textbf{Recall} & \textbf{EM} & \textbf{F1} \\
\midrule
\multirow{3}{*}{DeepSeek}
& No  & 21.2 & 24.6 & 36.8 & 48.7 \\
& Yes & 29.6 & 32.1 & 42.4 & 54.2 \\
& $\Delta$ & +8.4 & +7.5 & +5.6 & +5.5 \\
\midrule
\multirow{3}{*}{Qwen3-32B}
& No  & 19.8 & 24.2 & 37.6 & 50.0 \\
& Yes & 24.3 & 26.3 & 39.2 & 52.0 \\
& $\Delta$ & +4.5 & +2.1 & +1.6 & +2.0 \\
\bottomrule
\end{tabular}
\label{tab:brief-teacher-effect}
\end{table}

As shown in Table~\ref{tab:brief-teacher-effect}, search-state briefs improve downstream students under both teachers. With DeepSeek, the brief yields gains of 8.4 accuracy and 7.5 recall points on BrowseComp-Plus and 5.6 EM and 5.5 F1 points on Bamboogle. Qwen3-32B shows the same trend, gaining 4.5 accuracy and 2.1 recall points on BrowseComp-Plus and 1.6 EM and 2.0 F1 points on Bamboogle.

Qualitative trajectory inspection suggests that the brief mainly improves target preservation during teacher assessment and correction. Without it, the teacher may reformulate queries around a locally relevant but incorrect entity, omit original constraints, or misjudge evidence sufficiency. Retaining the original target, key constraints, and missing information provides a stable task-level anchor for student-specific corrections and reduces reinforcement of plausible but displaced directions. Although gains vary across teachers, their consistent direction indicates that the benefits of brief-guided assessment and correction are not teacher-specific.

\subsection{Role of Brief-Guided Assessment and Outcome Signals}
\label{sec:pair-construction-strategies}

We examine how preference-pair construction affects downstream performance. \textit{All-State Construction} retains every intermediate state. \textit{State-Only Selection} uses brief-guided assessment without rollout outcomes, whereas \textit{Outcome-Only Selection} considers only student \textsc{Answer} states and retains or corrects them based on rollout success. Our strategy combines both signals to construct informative pairs while avoiding supervision from states without a reliable local contrast.

\begin{figure}[t]
\centering
\includegraphics[width=\linewidth]{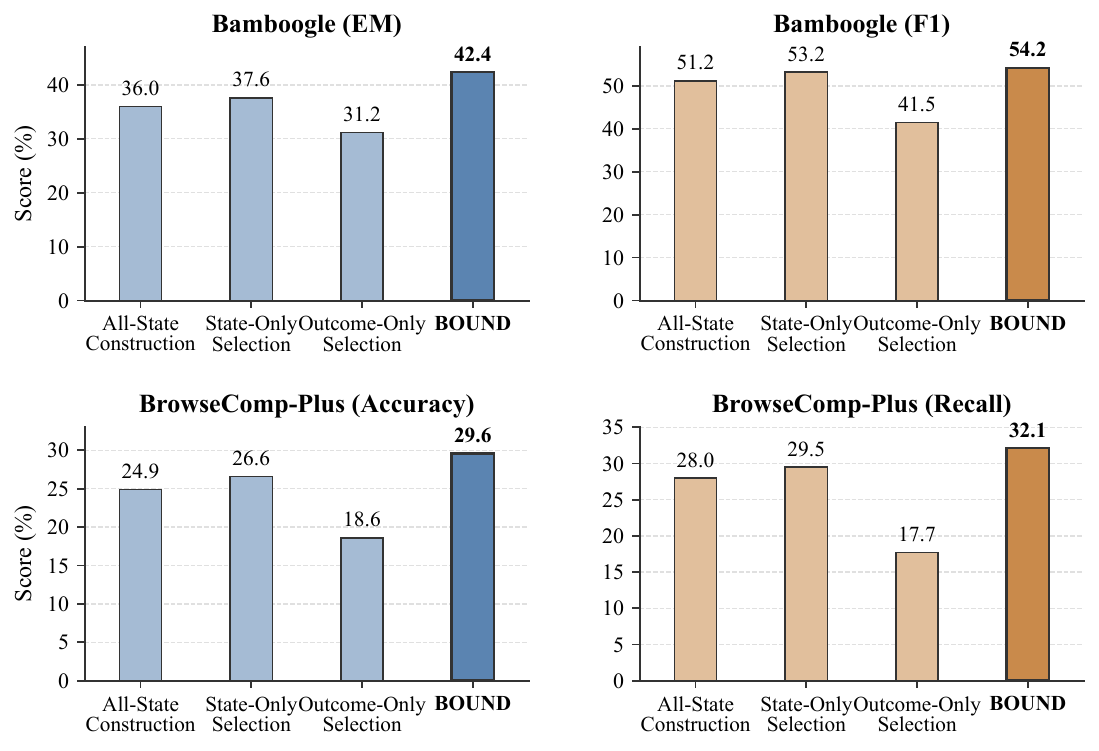}
\caption{Effect of preference-pair construction strategies. Higher is better.}
\Description{Comparison of four preference-pair construction strategies on Bamboogle EM and F1 and BrowseComp-Plus accuracy and recall. The combined strategy performs best across all metrics.}
\label{fig:pair-strategy}
\end{figure}

As shown in Figure~\ref{fig:pair-strategy}, our strategy performs best across all four metrics, supporting the complementarity of brief-guided assessment and rollout outcomes. All-State Construction may introduce weak or redundant supervision, while State-Only Selection lacks outcome information and Outcome-Only Selection cannot correct deviations before the final \textsc{Answer}. Combining both signals yields more locally grounded, outcome-informed corrective preferences.

\section{Conclusion}

We introduced BOUND, a brief-guided corrective preference distillation framework for multi-step information seeking. BOUND combines task-anchored search-state briefs, used during preference construction to preserve the original target and key constraints, with rollout outcomes to construct validated local preferences over student-induced states. It corrects erroneous continuations in unsuccessful rollouts while preserving supported answers in successful ones. Across multi-hop QA and deep search benchmarks, BOUND outperforms Trajectory SFT and a trained search baseline. In future work, we plan to evaluate BOUND with additional teachers and model scales.

\bibliographystyle{ACM-Reference-Format}
\bibliography{software}

\clearpage
\appendix
\section{Additional Results across 4B Backbones}
\label{app:qwen35-results}

To test whether BOUND depends on the Qwen3-4B-Instruct-2507 student, we additionally apply the full pipeline to Qwen3.5-4B~\cite{qwen35blog} and evaluate it on BrowseComp-Plus. We do not reuse Qwen3-4B preference data, ensuring that supervision remains specific to the stronger student's own failures. Qwen3.5-4B generates its own search rollouts, from which we independently construct validated state-matched preference pairs over student-induced decision-time states using the same brief-guided procedure.

\begin{table}[H]
\centering
\small
\caption{BrowseComp-Plus results across different 4B backbones. Each BOUND model is trained on preference pairs constructed from its own student rollouts.}
\label{tab:qwen35-bcplus}
\begin{tabular}{llcc}
\toprule
Backbone & Training & Accuracy (\%) & Recall (\%) \\
\midrule
\multirow{3}{*}{Qwen3-4B-Instruct}
& Before BOUND & 14.9 & 14.0 \\
& + BOUND & \textbf{29.6} & \textbf{32.1} \\
& $\Delta$ & +14.7 & +18.1 \\
\midrule
\multirow{3}{*}{Qwen3.5-4B}
& Before BOUND & 37.4 & 49.3 \\
& + BOUND & \textbf{44.0} & \textbf{53.4} \\
& $\Delta$ & +6.6 & +4.1 \\
\bottomrule
\end{tabular}
\end{table}

As shown in Table~\ref{tab:qwen35-bcplus}, BOUND improves both judge accuracy and retrieval recall across the two backbones. On the stronger Qwen3.5-4B student, accuracy increases by 6.6 percentage points, compared with a 4.1-point gain in recall. This larger accuracy improvement suggests that BOUND does more than retrieve additional relevant evidence: its student-specific corrections also reduce the influence of unsupported search directions and distracting context, helping the model use the available evidence more effectively. This further supports transfer across distinct rollout distributions and student-specific search-control errors.

\section{Manual Inspection of Search-State Briefs and Teacher Assessment}
\label{app:brief-quality}

We manually examine both the faithfulness of the search-state briefs and their effect on teacher-side assessment and correction. We randomly sample 50 intermediate states from those used for preference construction. Each state contains the original question, the history of preceding actions and their parameters, the active evidence context, and the student's original continuation.

\paratitle{Brief faithfulness.}
Each search-state brief records the Original Search Target, Key Constraints, Confirmed Evidence, Missing Information, and Drift Status. We manually inspect each field against the corresponding trajectory prefix. A field is considered valid only if it is consistent with the student-visible state and contains no major omission or unsupported information.

\paratitle{Paired assessment and correction inspection.}
To assess the brief's contribution, we evaluate the same teacher on 50 states under matched conditions. In the \emph{with-brief} condition, the teacher receives the student-visible state, structured brief, and original continuation; in the \emph{without-brief} condition, only the brief is omitted, with all other inputs and settings unchanged. Using identical decision-time states and student continuations isolates the effect of providing the brief.

We manually assign a reference drift status to each state before comparing the teacher outputs. Among the 50 inspected states, 13 are judged to be at risk of or already affected by drift, while the remaining 37 are aligned with the original search target and key constraints. For this comparison, at-risk and drifted states are jointly treated as drift-positive. We then inspect whether the teacher correctly assesses the state's drift status and whether its generated chosen continuation remains aligned with the original search target and key constraints. A chosen continuation is considered target-aligned only if it preserves or restores the intended target and constraints without continuing along a displaced direction. This criterion evaluates task-level alignment rather than surface wording and requires the continuation to remain grounded in information available at the original decision point.

\begin{table}[t]
\centering
\small
\caption{Manual inspection of search-state briefs and their effect on teacher assessment and correction.}
\label{tab:brief-quality}
\begin{tabular}{lcc}
\toprule
\multicolumn{3}{l}{\textbf{(a) Faithfulness of search-state brief fields}} \\
\midrule
Brief Field & Valid Cases & Ratio \\
\midrule
Original Search Target & 50/50 & 100.0\% \\
Key Constraints & 50/50 & 100.0\% \\
Confirmed Evidence & 49/50 & 98.0\% \\
Missing Information & 47/50 & 94.0\% \\
Drift Status & 47/50 & 94.0\% \\
\bottomrule
\end{tabular}
\begin{tabular}{lcc}
\toprule
\multicolumn{3}{l}{\textbf{(b) Assessment and correction quality}} \\
\midrule
Metric & With Brief & Without Brief \\
\midrule
Drift-status accuracy & 47/50 (94.0\%) & 45/50 (90.0\%) \\
Drift precision & 11/12 (91.7\%) & 12/16 (75.0\%) \\
Drift recall & 11/13 (84.6\%) & 12/13 (92.3\%) \\
False drift positives & 1/37 (2.7\%) & 4/37 (10.8\%) \\
Target-aligned chosen & 49/50 (98.0\%) & 46/50 (92.0\%) \\
\bottomrule
\end{tabular}
\end{table}

As shown in Table~\ref{tab:brief-quality}(a), all inspected briefs preserve the Original Search Target and Key Constraints, while 49 of the 50 briefs accurately summarize the Confirmed Evidence. Most errors occur in Missing Information and Drift Status, which require interpreting unresolved evidence and search direction rather than extracting explicit facts.

Table~\ref{tab:brief-quality}(b) further shows that the brief improves teacher-side assessment. With the brief, drift-status accuracy increases from 45/50 to 47/50, false drift positives decrease from four to one, and drift precision increases from 75.0 to 91.7. Drift recall decreases slightly from 92.3 to 84.6, indicating a more conservative assessment that avoids unnecessary correction of aligned states.

The target-alignment rate of the generated chosen continuation also increases from 92.0 to 98.0. Because the brief contains no future observations or gold answers, these gains reflect structured state abstraction rather than additional supervision, helping the teacher remain anchored to the original target, constraints, and unresolved evidence needs.

Although limited to 50 states, this analysis complements the downstream results by showing that the search-state brief is largely faithful to the student-visible state and improves the reliability and target alignment of teacher-side preference construction.

\section{Training Question Sources and Auxiliary Question Construction}
\label{app:synthetic-data}

\subsection{Training Question Sources}

We construct the corrective preference data from 983 distinct multi-hop questions. Using a fixed random seed of 42, we first sample 300 candidate questions from the training splits of HotpotQA~\citep{yang2018hotpotqa} and MuSiQue~\citep{trivedi2022musique}. We apply the same screening criterion to both sources, excluding questions that do not require substantive multi-step evidence acquisition under our search setting, such as those reducible to direct lookup or single-evidence answering. This procedure retains 274 existing questions, including 85 from HotpotQA and 189 from MuSiQue. Separately, applying the graph-based construction and validation procedure described in Section~\ref{app:question-construction} to the same English Wikipedia dump yields 709 auxiliary synthetic questions. We retain all questions that pass their respective procedures rather than subsampling either source to impose a predefined mixture ratio, resulting in 983 training questions in total. No evaluation questions from any benchmark are used during preference construction.

The existing questions provide naturally occurring relation patterns and linguistic variation, while the synthetic questions supplement them with convergent, multi-path evidence structures that are less frequent in standard multi-hop QA datasets. The two sources therefore contribute complementary question structures while sharing the same downstream supervision procedure. Synthetic construction supplies training questions rather than preference labels: questions from all sources are processed using the same student-rollout and corrective preference construction pipeline described in Section~\ref{sec:preference-construction}. Source metadata is retained in the construction records for auditing but is not included in the student-visible input.

\subsection{Auxiliary Synthetic Question Construction}
\label{app:question-construction}
To supplement existing multi-hop QA questions, we construct synthetic questions from the March 1, 2022 English Wikipedia dump. A local evidence graph identifies answer entities supported through multiple evidence paths. The teacher extracts grounded graph edges and verbalizes selected evidence structures rather than generating questions from unconstrained topics or evaluation examples.

\paratitle{Seed Sampling and Evidence-Graph Construction.}
We randomly sample Wikipedia documents as graph seeds using a fixed random seed. We filter list, disambiguation, template, category, timeline, outline, and year pages, and remove documents outside the specified length range. Each retained seed initializes graph exploration and need not correspond to the final answer.

We perform breadth-first expansion from each retained seed to a depth of three to four. For each expanded page, the teacher extracts neighboring named entities, relations connecting the current subject to those entities, and exact supporting spans. An edge is retained only if its supporting span occurs verbatim in the page and its target entity resolves to a document in the local corpus. Each retained edge is associated with its supporting passage and document identifier.

\paratitle{Convergent Candidates and Question Generation.}
We enumerate seed-to-entity paths and identify entities supported by multiple evidence paths. Selected paths cannot reuse intermediate nodes and must traverse distinct document sequences. A candidate answer must be supported by at least two such paths and three distinct evidence documents. We retain at most four paths per candidate.

Given a candidate and its evidence paths, the teacher generates one question whose constraints are derived only from the supplied evidence. Each path must contribute a distinct constraint, unsupported facts are prohibited, and the answer entity cannot be named directly. The teacher receives no evaluation questions, answers, or annotations. Each generated record stores the question, answer, constraints, evidence paths, supporting documents, and graph provenance.

\paratitle{Question Validation and Decontamination.}
We apply rule-based and teacher-based validation. The rule-based stage rejects malformed questions, questions outside the permitted length range, questions with fewer than two distinct constraints, duplicate questions, and questions revealing the normalized answer string. We also reject a question if the answer page appears among the top-5 results returned by the hybrid retriever used in our experiments.

The remaining questions undergo a single-pass grounding check. A question is retained only if all constraints are supported, the answer cannot be derived from a single document, no answer leakage is present, and at least three required document titles appear in the supplied evidence or retrieved results. The pipeline performs no scalar difficulty scoring, iterative prompt repair, or post-generation rewriting. Accepted and rejected questions are stored separately with their evidence and verification records.

We additionally audited the retained synthetic questions against all questions from the seven evaluation benchmarks. We found no exact or normalized duplicates and no complete benchmark question in the chosen, rejected, history, or context fields of the 1,670 preference pairs derived from synthetic questions. We manually reviewed stratified samples and the highest-scoring cases under complementary lexical and semantic similarity measures. The identified cases shared only broad topics or generic phrasing, while their key entities, constraints, solution paths, and answers remained distinct.

\section{Additional Details of Brief-Guided Corrective Preference Construction}
\label{app:preference-construction}

This appendix specifies the decision-time information used during preference construction, brief-guided assessment and outcome conditioning, teacher-side construction interfaces, pair construction and validation, student-facing serialization, and rerouting execution semantics. The search-state brief, assessment record, rollout outcome, and other construction-side information are used only to construct preference data and are excluded from preference optimization and inference inputs.

The implementation uses auxiliary generation tags to route candidate construction under the fixed search-agent interface. These tags are construction-time variables rather than an exhaustive taxonomy of search failures, and they are neither optimized nor exposed to the student. A decision point constitutes a \emph{search-control boundary} only when alternative continuations under the same student-visible state imply different local search-control decisions and the task and decision-time evidence support a reliable preference. A validated state-matched pair operationalizes such a boundary; no auxiliary tag identifies one independently.

\subsection{Decision-Time Assessment and Outcome Conditioning}
\label{app:retention-validation}

For each decision-time state $s_t=(q,H_t,C_t)$, the teacher first constructs a search-state brief:
\[
b_t=g_{\mathrm{B}}(s_t).
\]
The brief generator receives only the original question $q$, preceding actions and their parameters $H_t$, and evidence available in $C_t$.

The teacher then assesses the student's original continuation:
\[
d_t
=
(z_t,\rho_t,e_t)
=
\operatorname{Assess}(s_t,b_t,y_t^0),
\]
where $z_t$ is an auxiliary generation tag, $\rho_t$ is a state-grounded justification, and $e_t\in\{0,1\}$ indicates whether the continuation contains a specific local error. The tag takes one of four values:
\[
z_t\in
\left\{
\begin{aligned}
&\texttt{evidence\_completion},\quad
\texttt{target\_maintenance},\\
&\texttt{answerability},\quad
\texttt{not\_selected}
\end{aligned}
\right\}.
\]

The error indicator and auxiliary tag serve different purposes. The indicator $e_t$ evaluates whether the complete student continuation contains a specific local error. The tag $z_t$ indicates whether the state supports a reliable local contrast and, when it does, routes construction toward acquiring unresolved evidence, restoring the original target or a missing constraint, or preferring an evidence-supported answer.

The value \texttt{not\_selected} means that no pair-construction route is authorized. It is not a no-error label: it may occur with $e_t=0$ when the continuation is locally appropriate, or with $e_t=1$ when an error is apparent but no reliable correction can be constructed. It is also used when multiple alternatives are similarly reasonable or the available information does not establish a meaningful local preference. Thus, $z_t$ is neither a boundary label nor a substitute for the error judgment.

The error indicator evaluates the continuation's thought, action, and parameter rather than only its action label. For example, $e_t$ may be true when a \textsc{Continue} query pursues an unsupported entity instead of the unresolved information in the state.

The assessment receives $s_t$, $b_t$, and $y_t^0$, but not the observation produced by $y_t^0$, any later action or observation, the rollout outcome, or a gold answer. The rollout outcome is consulted only after assessment.

The operator in Algorithm~\ref{alg:preference-construction} returns the assessment record and an optional correction:
\[
(d_t,\widetilde{y}_t^{\mathrm{T}})
=
\operatorname{AssessAndCorrect}(s_t,b_t,y_t^0).
\]
When $e_t=1$ and $z_t\neq\texttt{not\_selected}$, the teacher attempts to generate
\[
\widetilde{y}_t^{\mathrm{T}}
=
\operatorname{Correct}(s_t,b_t,d_t,y_t^0).
\]
Otherwise, $\widetilde{y}_t^{\mathrm{T}}=\varnothing$. Any generated correction must still pass validation.

Let $r\in\{0,1\}$ denote the rollout outcome. For an unsuccessful rollout, BOUND constructs a corrective contrast when
\[
r=0
\;\land\;
e_t
\;\land\;
z_t\neq\texttt{not\_selected}
\;\land\;
\widetilde{y}_t^{\mathrm{T}}\neq\varnothing.
\]
The correction becomes the chosen continuation, and the student's original continuation becomes the rejected continuation. This condition neither treats every decision in an unsuccessful rollout as erroneous nor retains errors without reliable corrections.

For a successful rollout, define
\[
\begin{aligned}
\operatorname{SupportedAnswer}(d_t,y_t^0)
&\Longleftrightarrow
z_t=\texttt{answerability}\\
&\qquad\land \neg e_t
\land y_t^0\text{ is an \textsc{Answer}}.
\end{aligned}
\]
BOUND uses the earliest supported answer step:
\[
t^*
=
\min\{t:\operatorname{SupportedAnswer}(d_t,y_t^0)\}.
\]
At this step, the supported student answer is preferred to a plausible but unnecessary retrieval continuation.

Each contrast contributes to the preference dataset only if it passes validation. A surviving state-matched pair operationalizes a search-control boundary because its continuations share the same student-visible state, imply different local decisions, and admit a reliable preference under the task and decision-time evidence.

\subsection{Pair Construction and Validation}
\label{app:pair-validation}

For a corrective contrast, the candidate pair is
\[
\left(y_t^+,y_t^-\right)
=
\left(\widetilde{y}_t^{\mathrm{T}},y_t^0\right).
\]
The correction is conditioned on the state, brief, assessment record, and original continuation, and directly addresses the identified error under the existing action interface. Using the student's original continuation as the rejected continuation targets behavior that actually occurred under the same state.

For the earliest supported answer in a successful rollout, the original student answer remains chosen:
\[
\left(y_{t^*}^+,y_{t^*}^-\right)
=
\left(
y_{t^*}^0,
\operatorname{UnnecessaryRetrieval}
(s_{t^*},b_{t^*},y_{t^*}^0)
\right).
\]
The rejected continuation is a plausible retrieval action that continues searching after the available evidence already supports the answer.

In both cases,
\[
x_t=s_t.
\]
The chosen and rejected continuations therefore share the same question, preceding action history, and available evidence. Neither continuation is executed during pair construction, and no future observation, regenerated trajectory suffix, or additional evidence is attached to either side.

Each generated continuation contains exactly three non-empty fields:
\[
y_t
=
\left(
\texttt{thought},
\texttt{action},
\texttt{parameter}
\right),
\]
where
\[
\texttt{action}
\in
\left\{
\textsc{Continue},
\textsc{Reroute},
\textsc{Answer}
\right\}.
\]
For a corrective pair, the chosen continuation must directly address the identified error. For a termination pair, the rejected continuation must use \textsc{Continue} or \textsc{Reroute}. The two continuations must not be identical, express equivalent local search intent, or imply the same local decision.

Validation removes malformed or multiline fields, unsupported action labels, empty parameters, parameters longer than 512 characters, protocol tags embedded in parameters, and teacher-only information. It also removes pairs that differ only in superficial wording, cannot be supported from the decision-time state, or contain inconsistent actions and parameters.

Neither continuation may expose the search-state brief, auxiliary tag, assessment justification, rollout outcome, future information, or environment-side passage-selection details. For a termination pair, the chosen answer must be supported by evidence already present in the state, and the rejected continuation must pursue a concrete but unnecessary retrieval subgoal rather than an unrelated topic.

Exact duplicate student-facing records are removed. If the same student-visible state receives incompatible preferred action labels across candidate records, the corresponding ambiguous records are excluded to avoid introducing contradictory supervision. For auditing and reproducibility, construction-side records retain rollout and step identifiers, the rollout outcome, search-state brief, assessment, generated continuations, and validation result separately from the final student-facing DPO artifact.

\subsection{Teacher-Side Construction Interfaces}
\label{app:teacher-prompts}

Preference construction produces four teacher-side records: the brief, a local assessment, a correction for an erroneous continuation, and an unnecessary-retrieval alternative for a supported answer. Their separation is an implementation choice rather than part of the learning objective. The final optimization instance contains only the student-visible state and the validated chosen and rejected continuations.

The brief generator observes only $s_t$. The local assessment receives $s_t$, $b_t$, and $y_t^0$. Correction generation additionally receives $d_t$, while unnecessary-retrieval generation receives the state, brief, and supported student \textsc{Answer} after the decision point has been selected. None of these interfaces receives a future observation, a regenerated trajectory suffix, the rollout outcome, or a gold answer.

We reproduce below the prompts whose literal wording specifies the structured brief and unnecessary-retrieval records. For local assessment and correction generation, we report the input restrictions, output schemas, abstention behavior, and generation requirements in text; exact templates appear in the released code. In the prompt wording, ``search history'' refers to the preceding actions and parameters represented by $H_t$.

\begin{promptbox}{Search-State Brief Prompt}
Summarize the decision-time search state. Use only the question, search history, and evidence provided at this state. Do not use future actions, future observations, or a gold answer.\par
Return JSON only with exactly these five fields:\par
\begingroup
\raggedright
\ttfamily
\{\\
\hspace*{1em}"Original Search Target": "<target asked for by the question>",\\
\hspace*{1em}"Key Constraints": "<constraints the answer must satisfy>",\\
\hspace*{1em}"Confirmed Evidence": "<question-relevant facts supported by the current evidence>",\\
\hspace*{1em}"Missing Information": "<information still needed, or None>",\\
\hspace*{1em}"Drift Status": "<aligned|at risk|drifted>: <short reason>"\\
\}\par
\endgroup
Evidence sufficiency must be represented through Confirmed Evidence and Missing Information. Drift Status should describe only whether the current search direction remains anchored to the original target.\par
Do not restate the full question or copy entire evidence passages. Do not add fields or instructions.
\end{promptbox}

\paratitle{Local decision assessment.}
Given $s_t$, $b_t$, and $y_t^0$, the teacher determines whether the continuation contains a specific local error and whether the state supports a reliable local contrast.

The assessment is serialized as
\[
\texttt{\{boundary, reason, student\_action\_is\_error\}},
\]
corresponding to $z_t$, $\rho_t$, and $e_t$. The field name \texttt{boundary} is retained for compatibility with the implementation schema, but its value functions only as an auxiliary generation tag and does not independently identify a search-control boundary.

The field \texttt{student\_action\_is\_error} evaluates the complete continuation. The \texttt{boundary} field is set to \texttt{evidence\_completion}, \texttt{target\_maintenance}, or \texttt{answerability} when the state supports the corresponding corrective or termination contrast; otherwise, it is set to \texttt{not\_selected}.

A \texttt{not\_selected} tag covers locally appropriate continuations, errors without reliable corrections, several similarly reasonable alternatives, and states without a meaningful local preference. Such cases yield neither a correction nor a preference pair.

The assessment uses only the decision-time state, brief, and original continuation. It excludes the resulting observation, future trajectory information, rollout outcome, and gold answer.

\paratitle{Student-specific correction generation.}
When $e_t=1$ and $z_t\neq\texttt{not\_selected}$, the teacher attempts to generate a correction conditioned on $s_t$, $b_t$, $d_t$, and $y_t^0$. The correction targets the identified error under the fixed student action interface. It may acquire unresolved evidence, restore the original target or a missing constraint, or answer when the available evidence is sufficient. No correction is generated when $z_t=\texttt{not\_selected}$.

The generated record contains exactly three fields:
\texttt{thought}, \texttt{action}, and \texttt{parameter}.
The first briefly explains the next decision, while the latter two specify the student action and its focused retrieval query or concise final answer. Retrieval actions must target a concrete unresolved information need, and an \textsc{Answer} must be supported by evidence already available in $s_t$.

Because generation is conditioned on the original continuation, the correction addresses the student's specific error rather than replacing it with a generic teacher continuation. It must remain expressible from the student-visible state and must not copy teacher-only information into the student-facing output.

\begin{promptbox}{Unnecessary Retrieval Prompt}
Construct one rejected local continuation for an answerability decision point.\par
The chosen continuation correctly stops and answers using the currently available evidence. Construct a plausible alternative that performs unnecessary additional retrieval instead.\par
The rejected continuation must:\par
\noindent\hangindent=1.2em\hangafter=1\texttt{-} be plausible under the student-visible state;\par
\noindent\hangindent=1.2em\hangafter=1\texttt{-} use either \texttt{Continue} or \texttt{Reroute};\par
\noindent\hangindent=1.2em\hangafter=1\texttt{-} pursue a concrete but unnecessary subgoal suggested by the current state;\par
\noindent\hangindent=1.2em\hangafter=1\texttt{-} avoid introducing a new entity, constraint, or information need;\par
\noindent\hangindent=1.2em\hangafter=1\texttt{-} remain grounded in the student-visible state; and\par
\noindent\hangindent=1.2em\hangafter=1\texttt{-} express a clear contrast with the chosen Answer continuation.\par
Return JSON only:\par
\texttt{\{}\\
\hspace*{1em}\texttt{"thought"}: \texttt{"<brief rationale that incorrectly claims further retrieval is needed>"},\\
\hspace*{1em}\texttt{"action"}: \texttt{"<Continue|Reroute>"},\\
\hspace*{1em}\texttt{"parameter"}: \texttt{"<focused but unnecessary retrieval query>"}\\
\texttt{\}}\par
Do not introduce future observations or teacher-only information.
\end{promptbox}

\subsection{Student-Facing Serialization and Reroute Execution}
\label{app:preference-serialization}

Each validated pair yields one preference instance
\[
\left(x_t,y_t^+,y_t^-\right),
\qquad
x_t=s_t.
\]
Teacher outputs are generated as JSON for parsing and validation and then rendered as
\[
\begin{aligned}
&\texttt{[Thought]: <decision rationale>}\\
&\texttt{[Action]: <Continue|Reroute|Answer>}\\
&\texttt{[Parameter]: <query or answer>}.
\end{aligned}
\]

The final DPO artifact stores $x_t$, $y_t^+$, and $y_t^-$ as \texttt{prompt}, \texttt{chosen}, and \texttt{rejected}. The prompt contains the same search-policy instruction and student-visible state used during inference, while the chosen and rejected fields each contain one assistant continuation in the tagged format above. The brief, auxiliary tag, assessment justification, rollout outcome, validation metadata, and all other construction-side fields are omitted.

The state transition associated with \textsc{Reroute} is an environment-side execution rule rather than part of preference construction. When a policy selects \textsc{Reroute}, the action and rerouting query are appended to the history. The environment then computes the E5-base-v2 cosine similarity between the rerouting query and each passage returned by the immediately preceding retrieval. Passages with similarity below 0.2 are removed. If all passages fall below this threshold, the highest-scoring passage is retained. All earlier history and evidence are preserved.

The rerouting query is then issued to the benchmark-specific retrieval backend, which returns a new top-$5$ passage batch. The new passages are combined with earlier evidence and retained passages to form the next student-visible state. For BrowseComp-Plus, Qwen3-Embedding-8B performs the new retrieval, while E5-base-v2 only scores passages from the preceding retrieval.

This transition requires no teacher call and is applied during rollout collection and evaluation of the initial search policy, Trajectory SFT, and BOUND. Passage filtering is disabled only in the query-only rerouting ablation. Algorithm~\ref{alg:preference-construction}, by contrast, operates on fixed student rollouts and constructs chosen and rejected continuations without executing either continuation or modifying their shared student-visible state.

The final training artifact contains 2,424 validated step-level preference instances constructed from 983 distinct training questions.

\end{document}